\documentclass{webofc}
\usepackage[varg]{txfonts}   
\begin{document}
\title{Muon bundles from the Universe}
%
%

\author{\firstname{P.} \lastname{Kankiewicz}\inst{1}\fnsep \and
        \firstname{M.} \lastname{Rybczy\'{n}ski}\inst{1}\fnsep\thanks{\email{maciej.rybczynski@ujk.edu.pl}} \and
        \firstname{Z.} \lastname{W\l odarczyk}\inst{1}\fnsep \and
        \firstname{G.} \lastname{Wilk}\inst{2}\fnsep
}

\institute{Institute of Physics, Jan Kochanowski University, 25-406 Kielce, Poland
\and
National Centre for Nuclear Research, Department of Fundamental Research, 00-681 Warsaw, Poland
          }

\abstract{%
Recently the CERN ALICE experiment, in its dedicated cosmic ray run, observed muon bundles 
of very high multiplicities, thereby confirming similar findings from the LEP era at CERN 
(in the CosmoLEP project). Significant evidence for anisotropy of arrival directions 
of the observed high multiplicity muonic bundles is found. Estimated directionality suggests 
their possible extragalactic provenance. We argue that muonic bundles of highest multiplicity 
are produced by strangelets, hypothetical stable lumps of strange quark matter infiltrating 
our Universe.  
}
\maketitle
\section{Introduction}
\label{intro}
This talk is based on~\cite{Kankiewicz:2016dha}, where the relevant details can be found.
Cosmic ray physics is our unique source of information on events in the energy range which will never be accessible in Earth-bound experiments~\cite{DR,L-SS}. This is why one of the most important aspects of their investigation is the understanding of the primary cosmic ray (CR) flux and its composition. In this respect the recent measurement performed by the ALICE experiment at CERN LHC in its dedicated cosmic ray run~\cite{ALICE:2015wfa,ALICE:2015wfa_suppl} is of great importance. A number of events with muon bundles of high multiplicity was registered in the so called Extensive Air Showers (EAS) produced by cosmic ray interactions in the upper atmosphere~\cite{ALICE:2015wfa}. A special emphasis has been given to the study of high multiplicity events containing more than $100$ reconstructed muons. Similar events have already been studied in the previous LEP experiments at CERN (in the so called CosmoLEP program) such as ALEPH, DELPHI and L3. 

In this talk we shall concentrate mainly on the first, high multiplicity events. To describe them we propose seriously to consider, for a moment, the possibility of the existence in the flux of incoming CR a component with very high atomic number, of the order of $A \sim 10^3$. In fact we propose to return to our old idea that muon bundles of extremely high multiplicities could be produced by \emph{strangelets}, hypothetical stable lumps of strange quark matter (SQM) infiltrating our Universe~\cite{Ryb}. Strangelets with such masses, much larger than the masses of ordinary nuclei, could easily produce extremely large groups of muons in collisions with the atmosphere.

\begin{figure}[h]
\centering
\includegraphics[width=9.5cm,clip]{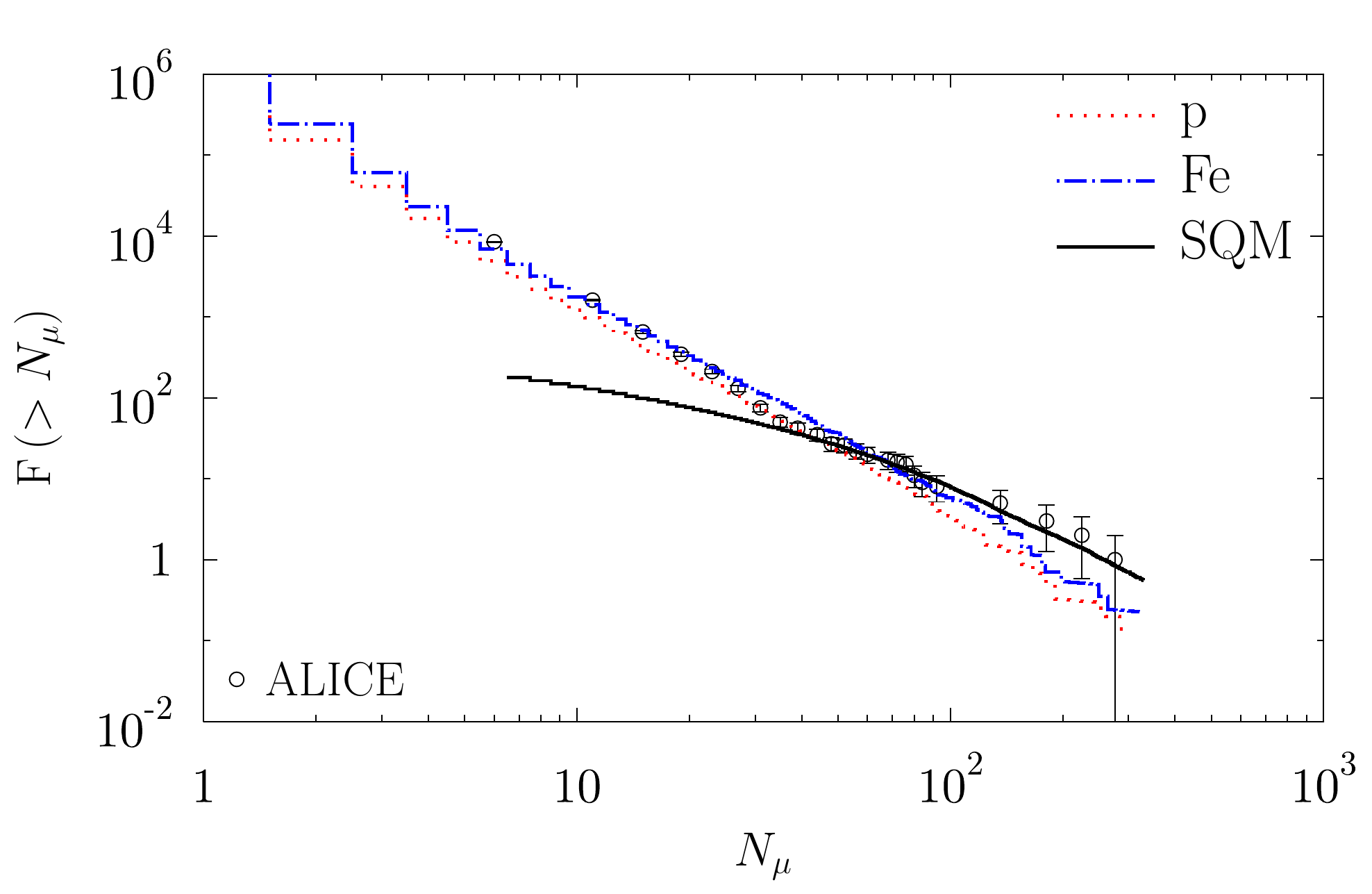}
\vspace{-0.25cm}
\caption{Integral multiplicity distribution of muons for the ALICE data (circles) published in~\cite{ALICE:2015wfa}. Monte Carlo simulations for primary protons (dotted line); iron nuclei (dashed dot line) and primary strangelets with mass A taken from the $A^{-7.5}$ distribution (full line) with abundance of the order of $2\cdot 10^{-5}$ of the total primary flux.}
\label{fig-ALICE}       
\end{figure}

\section{Searches for Strange Quark Matter in cosmic rays}
\label{sec-1}
The idea of SQM originated some time ago in Ref.~\cite{Witten}, but it remains alive, cf. \cite{S-MGA}. In short, the basic, commonly accepted view is that that SQM (understood as a combination of roughly an equal number of up, down and strange quarks) might be the true ground state of quantum chromodynamics (QCD). It is therefore reasonable to expect that it exists in some form in the Universe and can be detected. This supposition resulted in a number of searches for strange stars and quark stars, in which such a form of matter would be dominant and which could therefore be a possible source of strangelets penetrating outer space~\cite{QS-Science}. 
The experimental data mentioned above lead to a flux of strangelets which follows the $A^{-7.5}$ behaviour, which in turn coincides with the behaviour of the abundance of normal nuclei in the Universe. Strange quark matter fills a vast gap in the distribution of all known forms of stable matter, cf. chart of nuclides presented in Ref.~\cite{Crawford:1994cn}. It places itself exactly between the heaviest atomic elements and neutron stars.
The estimated flux of strangelets is also consistent with the astrophysical limits and with the upper limits given experimentally~\cite{Sahnoun:2008mr}. 

\section{High multiplicity muon bundles from SQM}
\label{sec-2}
We start with a short reminder of our proposed possible scenario of propagation of strangelets through the atmosphere \cite{Wilk:1996jpg}. The apparent contradiction between the large initial size of the incoming strangelets  and their required strong penetrability in the atmosphere can be resolved by assuming that strangelets reaching deeply into the atmosphere are formed from the original large strangelet which loses mass in many successive interactions with air nuclei when penetrating the atmosphere. To provide numerical estimate we limit ourselves to the two most extreme pictures of the collision of a strangelet of mass number $A$ with an air nucleus target of mass number $A_t$: $(i)$ All quarks of $A_t$ which are located in the geometrical intersection of the two colliding projectiles are involved and one assumes that each quark from the target interacts with only one quark from the strangelet; $(ii)$ All quarks from both nuclei which are in their geometrical intersection region participate in the collision. In the first case, during the interaction up to ($3A_t$) quarks from the strangelet could be used up and its mass could drop to a value of $A - A_t$, at most. The second case is an analogue of the so called tube model used occasionally in nuclear collisions. After the collision the atomic mass of the strangelet diminishes to a value equal to $A -  A^{1/3}\cdot A_t^{2/3}$.  While this is a rather extreme variant, it is still useful in providing  an estimate of the maximum possible destruction of the quarks in the strangelet.

In this study we have used suitably modified SHOWERSIM~\cite{SHOWERSIM84} modular software. We performed Monte Carlo simulations for primary nuclei composed of protons and iron nuclei and for primary strangelets with mass $A$ taken from the $A^{-7.5}$ distribution. The results of our simulations are shown in Fig.~\ref{fig-ALICE}. Note that whereas the lower and medium multiplicities can be reproduced by the ordinary nuclei, the extremely large groups of muons can be described only by allowing (a relatively minute, of the order of $10^{-5}$ of the total primary flux) admixture of SQM of the same total energy.

\section{Anisotropy of arrival directions of strangelets}
\label{sec-3}
The ALICE data also turns out to be very valuable for another reason. Namely, they show the angular distribution of the muon events in the spherical reference frame with zenith angle ($\theta$) and azimuth angle ($\Phi$).

Taking into account the orientation of the axes, these coordinates can be transformed into a horizontal reference frame: $(\Phi,\theta)\to(Az,h)$, where $Az$ is the Azimuth used in navigation/astronomy (in this particular case: $Az=\Phi+\alpha_{0}$), and  $h$ is the elevation above the horizon (where $h=90^{\circ}-\theta$). Finally, we can transform the coordinates from the horizontal to the equatorial reference frame: $(Az,h)\to(\alpha,\delta)$. Such a system of spherical coordinates is often used in astronomy and is expressed by right ascension ($\alpha$) and declination ($\delta$). 

\begin{figure}[h]
\centering
\includegraphics[width=10cm,clip]{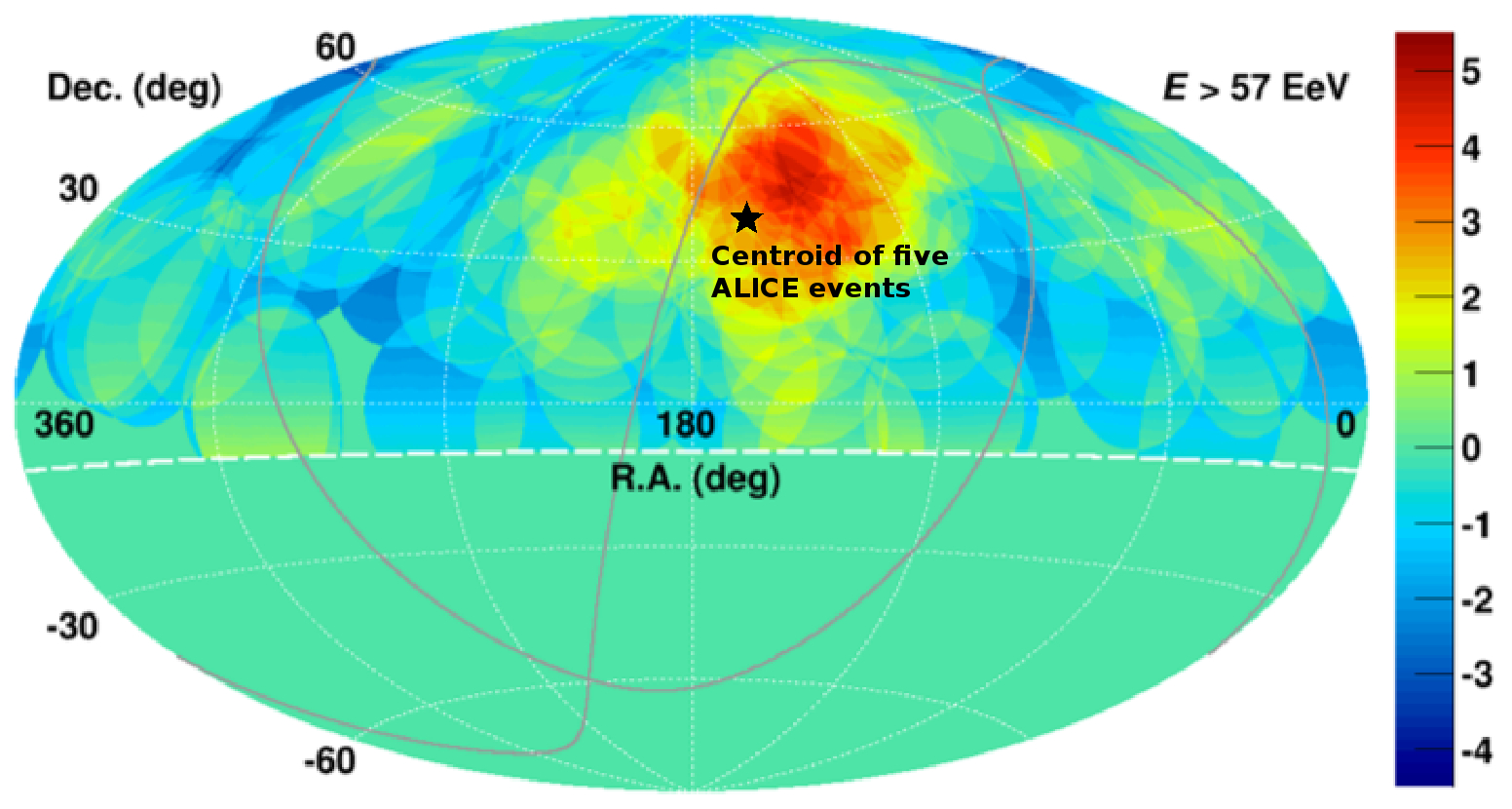}
\caption{Aitoff projection of the UHECR map in equatorial coordinates taken from Telescope Array Collaboration data~\cite{Abbasi:2014lda}. Location of the centroid of five ALICE events is marked with a star.}
\label{fig-centroid}       
\end{figure}

Using the usual spherical triangle formulas or rotation matrices one can convert between the horizontal and equatorial coordinates. For the five high-multiplicity muon events we obtained the celestial equatorial coordinates $(\alpha_{2000},\delta_{2000})$ and the estimated limits of their positional errors are $\sim 10^{\circ}$~\cite{Kankiewicz:2016dha}. The centroid (average location) of the five considered events is marked by a star in Fig. \ref{fig-centroid}. The coordinates of this centroid are: $\alpha_{centr} = 11~h~15~m$ , $\delta_{centr} = 39^{\circ}$ $29^{\prime}$.

We can conclude that the distribution of the directions of these five events suggests their possible extragalactic provenance. Most of them are located in the vicinity of the galactic north pole, far from the galactic plane. If we temporarily assume that all these events have the same source, the most probable source would be the blazar Mrk 421 which is located close to the centroid of the five events. It is one of the brightest blazars known, with major outbursts, active at all wavebands \cite{Horan:2004}. However, there are also many other known extragalactic high energy sources located close to the coordinates of our events \cite{Horan:2004}, \cite{Turley:2016ver} and, at the moment, they cannot be excluded as potential sources. 

\section{Conclusions}
\label{sec-4}
There is an ongoing discussion concerning the possibility of the existence in the Universe of some stable forms of  SQM and the feasibility of its detection. Whereas until now we do not have fully convincing observations of SQM,  nevertheless there are numerous indications of the possible existence of strange stars, or even of the existence of quark stars. Strangelets could originate, for example, from violent collisions between such objects. The recent results of ALICE reporting the existence of bundles of high-multiplicity muons could serve, notwithstanding the weakness of this signal, as a direct (almost) signal of stable lumps of SQM called strangelets arriving at the Earth from outer space. As a result, we have found that whereas the low multiplicities of muon groups measured by the CERN ALICE experiment favour light nuclei as primaries and the medium multiplicities show behaviour specific for heavier primaries, the muon groups (bundles) of really high multiplicities (of the order of ~$\sim 100$) apparently cannot be described by the common interaction models. We have shown here that the situation can be rescued by allowing for a relatively small (of the order of $10^{-5}$ of the total primary flux) admixture of strangelets of the same total energy. Our estimation of their flux does not contradict the results obtained recently by the SLIM Collaboration~\cite{Sahnoun:2008mr}. The ALICE data allowed us to estimate the directionality of the five events of the highest muon multiplicities, and to identify their most probable extragalactic source(s). 

\section*{Acknowledgments}
This research was supported by the Polish National Science Centre, grant 2016/23/B/ST2/00692 (MR) and contract 2016/22/M/ST2/00176 (GW).
%

\begin{thebibliography}{}
%
%

\bibitem{Kankiewicz:2016dha} 
  P.~Kankiewicz, M.~Rybczynski, Z.~Włodarczyk and G.~Wilk,
  Astrophys.\ J.\  {\bf 839}, no. 1, 31 (2017).

\bibitem{DR}
 A.~Dar and A.~De~R\,ujula,
Phys. Rep. {\bf 466} 179 (2008).

\bibitem{L-SS}
A.~Letessier-Selvon and T.~Stanev,
Rev. Mod. Phys. {\bf 83} 907 (2011).

\bibitem{ALICE:2015wfa}
J.~Adam {\it et al.} [ALICE Collaboration],
J. Cosm. Astrop. Phys. {\bf 01} 032 (2016).

\bibitem{ALICE:2015wfa_suppl}
J.~Adam {\it et al.} [ALICE Collaboration],
ALICE-PUBLIC-2016-003.

\bibitem{Ryb} M. Rybczy\'nski, Z. W\l odarczyk, G. Wilk,
Nucl. Phys. B (Proc. Suppl.) {\bf 97}, 85 (2001). 

\bibitem{Witten}
 E.~Witten,
Phys. Rev. D {\bf 30} 272 (1984).

\bibitem{S-MGA}
 M.~G.~Alford,
Nucl. Phys. A {\bf 830} 385c (2009).

\bibitem{QS-Science}
 K.~S.~Cheng, Z.~G.~Dai, D.~M.~Wei and T.~Lu,
Science {\bf 280} 407 (1998).

\bibitem{Crawford:1994cn}
  H.~J.~Crawford and C.~H.~Greiner,
Sci. Am.  {\bf 270} 58 (1994). 

\bibitem{Sahnoun:2008mr}
  Z.~Sahnoun {\it et al.},
Radiat. Meas.  {\bf 44} 894(2009).

\bibitem{Wilk:1996jpg}
G.~Wilk and Z.~Wlodarczyk, 
J. Phys. G {\bf 23} 2057 (1997). 

\bibitem{SHOWERSIM84}
  A.~Wrotniak, SHOWERSIM/84, University of Maryland Preprint, 85 (1984).

\bibitem{Horan:2004}
D.~Horan and T.~C.~Weekes,
New Astr. Rev. {\bf 48} 527 (2004).

\bibitem{Turley:2016ver} 
  C.~F.~Turley {\it et al.} [Astrophysical Multimessenger Observatory Network) Collaboration],
  Astrophys.\ J.\  {\bf 833}, no. 1, 117 (2016).

\bibitem{Abbasi:2014lda} 
  R.~U.~Abbasi {\it et al.} [Telescope Array Collaboration],
  Astrophys.\ J.\  {\bf 790}, L21 (2014).

\end{thebibliography}
%
%

\end{document}